\documentclass[prb,showpacs,floatfix,twocolumn,superscriptaddress]{revtex4}
\usepackage{graphicx}
\usepackage{dcolumn}

\begin{document}

\newcommand*{\cm}{cm$^{-1}$\,}
\newcommand*{\cis}{CuIr$_2$S$_4$\,}
%
\title{Optical study of the metal-insulator transition in CuIr$_2$S$_4$ crystals}
%
%

\author{N. L. Wang}
\email{nlwang@aphy.iphy.ac.cn}%
\affiliation{Institute of Physics and Center for Condensed Matter
Physics, Chinese Academy of Sciences, P.~O.~Box 603, Beijing
100080, P.~R.~China}
\author{G. H. Cao}
\affiliation{Department of Physics, Zhejiang University, Hangzhou
310027, P.~R.~China}
\author{P. Zheng}
\author{G. Li}
\author{Z. Fang}
\affiliation{Institute of Physics and Center for Condensed Matter
Physics, Chinese Academy of Sciences, P.~O.~Box 603, Beijing
100080, P.~R.~China}
\author{T. Xiang}
\affiliation{Institute of Theoretical Physics and
Interdisciplinary Center of Theoretical Studies, Chinese Academy
of Sciences, Beijing 100080, P.~R.~China}
\author{H. Kitazawa}
\author{T. Matsumoto}
\affiliation{National Institute for Materials Science, Sengen
1-2-1, Tsukuba, Ibaraki 305-0047, Japan}
%
%
%
\begin{abstract}
We present measurements of the optical spectra on single crystals
of spinel-type compound \cis. This material undergoes a sharp
metal-insulator transition at 230 K. Upon entering the insulating
state, the optical conductivity shows an abrupt spectral weight
transfer and an optical excitation gap opens. In the metallic
phase, Drude components in low frequencies and an interband
transition peak at $\sim 2 eV$ are observed. In the insulating
phase, a new peak emerges around $0.5 eV$. This peak is attributed
to the transition of electrons from the occupied Ir$^{3+}$
$t_{2g}$ state to upper Ir$^{4+}$ $t_{2g}$ subband resulting from
the dimerization of Ir$^{4+}$ ions in association with the
simultaneous formations of Ir$^{3+}$ and Ir$^{4+}$ octamers as
recently revealed by the x-ray diffraction experiment. Our
experiments indicate that the band structure is reconstructed in
the insulating phase due to the sudden structural transition.
\end{abstract}

\pacs{72.80.Ga, 78.20.Ci, 71.30.+h, 78.30.-j}

\maketitle

%
%
Spinel type compound \cis has recently attracted much attention
for its intriguing first-order metal-insulator transition (MIT) at
T$_{MI}$$\sim$ 230 K
\cite{Nagata1,Furubayashi,Oda,Matsuno,Nagata2,Matsumoto,Suzuki,Burkov,Hayashi,Cao1,Radaelli,Ishibashi,Furubayashi2,Cao2,Croft}.
The transition is characterized by a sudden increase of the
electrical resistivity, a disappearance of Pauli paramagnetism, a
hysteresis loop in resistivity $\rho$ and magnetic susceptibility
$\chi$, and a lowering of structure symmetry. Above the MIT
temperature, \cis has a normal cubic spinel structure, in which
the Cu ions (A sites) are tetrahedrally coordinated and the Ir
ions (B sites) are octahedrally coordinated by sulfur ions. Upon
entering the low temperature insulating phase, a structural
deformation occurs, lowering the lattice symmetry to triclinic
\cite{Radaelli}.

The structure and the MIT in \cis is reminiscent of a classic
spinel compound--the magnetite Fe$_3$O$_4$, which also exhibits an
abrupt MIT at about 120 K, called the Verway transition
\cite{Verwey}. The Fe$_3$O$_4$ undergoes a ferrimagnetic
transition at a much higher temperature (858 K). Below this
temperature, the magnetic moments of the Fe ions are
ferrimagnetically ordered, but the A sites [Fe$^{3+}$
(t$_{2g}^3e_g^2$, s=5/2)] and B sites [Fe$^{2+}$ ($t_{2g}^4e_g^2$,
s=2) and Fe$^{3+}$ (s=5/2)] have opposite spin directions. The
Verway transition has been interpreted as a charge ordering
transition of Fe$^{2+}$ and Fe$^{3+}$ on the B sites in alternate
(001) planes \cite{note1}.

Naturally, it is considered that the MIT in \cis is similar to the
Verway transition in Fe$_3$O$_4$. Since the band structure
calculation and the photoemission experiments revealed that the
valence state of Cu is Cu$^{1+}$ \cite{Oda,Matsuno}, it is
believed that the ionic configuration of
Cu$^{1+}$Ir$^{3+}$Ir$^{4+}$S$_4^{2-}$ is realized in the low-T
insulating phase, and the ordering of Ir$^{3+}$ and Ir$^{4+}$ ions
is a plausible origin of the MIT \cite{Matsuno,Cao1}. Indeed,
recent high-resolution synchrotron X-ray powder diffraction
experiment revealed a peculiar form of charge ordering which
consists of alternating arrangement of isomorphic octamers or
clusters of Ir$_8^{3+}$S$_{24}$ and Ir$_8^{4+}$S$_{24}$ (as
isovalent bi-capped hexagonal rings) together with
spin-dimerizations between Ir$^{4+}$ ions \cite{Radaelli}. The
charge-ordering pattern is much more complicated than Fe$_3$O$_4$
as well as any other previously known charge-ordered structures
which are typically based on stripes, slabs or chequerboard
patterns. The simultaneous charge-ordering and spin-dimerization
transition is a rare phenomenon in three-dimensional
compounds\cite{Radaelli}. Therefore, it is very interesting to
further explore how the electronic structures change in the MIT.
This work presents a detailed infrared spectroscopy study on
single crystal samples. It provides important information about
low-lying excitations across the transition.

Single crystals of \cis were grown from the bismuth solution
\cite{Matsumoto2}. First, single phase \cis powders were
synthesized by solid-state reaction in sealed quartz tube using
high purity (better than 4N) powders of elements Cu, Ir, and S as
starting materials. Then, \cis and metal bismuth (6N) in the molar
ratio of 1:100 were sealed in an evacuated quartz ampoule. The
ampoule was heated to 1273 K, and holding for two days. Crystals
of \cis were grown by cooling at 4 K/hour down to 773 K.
Typically, the crystals have triangular shape of surface with edge
length about 0.4 mm. The near-normal incidence reflectance spectra
were measured by using a Bruker 66v/S spectrometer in the
frequency range from 100 \cm to 28000 \cm. The sample was mounted
on an optically black cone in a cold-finger flow cryostat. An
\textit{in situ} overcoating technique was employed for
reflectance measurement \cite{Homes}, which enables us to get
reliable data on small-size samples. The spectra above 500 \cm was
collected on one single crystal, while the data in the
far-infrared regime was obtained on mosaic crystal samples. The
optical conductivity spectra were obtained from a Kramers-Kronig
transformation of R($\omega$). We use Hagen-Rubens' relation for
the low frequency extrapolation, and a constant extrapolation to
80000 \cm followed by a well-known function of $\omega^{-4}$ in
the higher-energy side.

\begin{figure}[t]
\centerline{\includegraphics[width=2.9in]{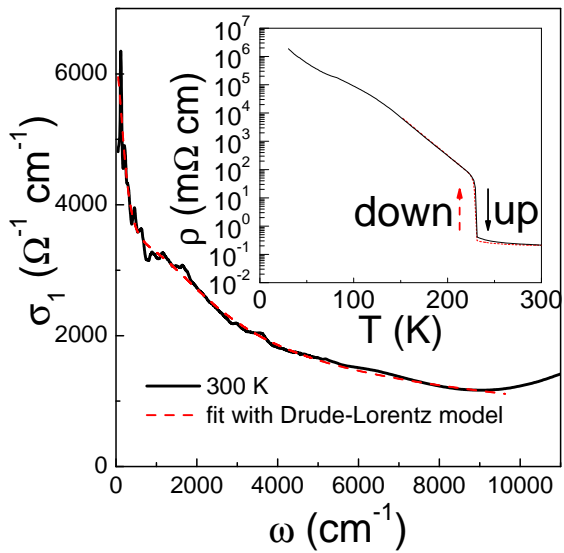}}%
\vspace*{-0.2cm}%
\caption{(Color online) The optical conductivity spectrum of \cis
at 300 K. Dashed curve is a fit to the data using two Drudes and
one Lorentz components. The inset shows T-dependence of dc resistivity
as the sample was cooled down and warmed up.}%
\label{1}
\end{figure}

Fig. 1 shows the room-T optical conductivity below 11000 \cm. The
dc resistivity as a function of temperature is shown in the inset.
There is a sharp metal-insulator transition at 230 K. The optical
spectrum shows an usual metallic response with a conductivity
minimum around 9000 \cm. The spectrum could be well fitted with
two Drude components and a Lorentz oscillator below the frequency
of minimum conductivity. The two Drude components, which result
from bands crossing the Fermi energy, have plasma frequencies and
scattering rates of $\omega_{p1}\approx$ 7000\cm,
$\Gamma_1\approx$ 250 \cm and $\omega_{p2}\approx$ 20000\cm,
$\Gamma_2\approx$ 2400 \cm, respectively. The Lorentz part has a
central frequency of 4000 \cm (0.5 eV).

Fig. 2 shows the reflectance and optical conductivity spectra at
different temperatures over broad frequencies. The spectra show
little change as temperature decreases from 300 K to 232 K, except
in the very low-frequency region. However, upon entering the
insulating phase, dramatic change occurs in optical spectra. The
low-$\omega$ spectral weight below 0.5 eV (4000 \cm) is severely
suppressed, resulting in the opening of an optical gap. The
missing spectral weight is transferred to higher energies, forming
a pronounced peak ($\alpha$) at 0.5 eV. In addition, another peak
(labelled as $\beta$) exists around 2 eV (16000 \cm). This peak is
also present in the metallic state in high temperatures at
slightly higher frequency. The strong suppression of the
low-energy spectral weight and the two-peaks ($\alpha$ and
$\beta$) structure are the most pronounced features below
T$_{MI}$.

\begin{figure}[t]
\centerline{\includegraphics[width=2.9in]{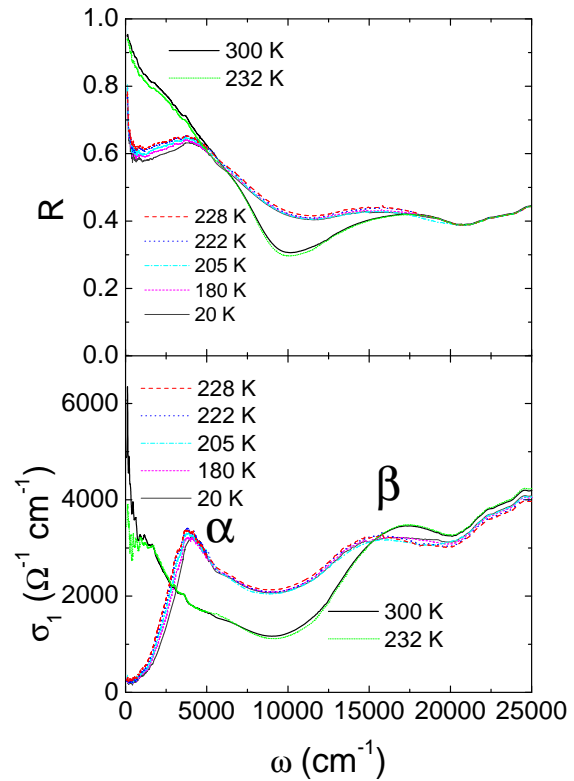}}%
\vspace*{-0.2cm}%
\caption{(Color online) Temperature dependence of the reflectivity
spectra (up panel) and optical conductivity spectra (bottom panel)
for a \cis
single crystal with $T_{MI}$=230 K. Abrupt spectral change occurs at $T_{MI}$.}%
\label{fig2}
\end{figure}

Understanding the above spectral change, which is the main task of
this work, is crucial for the understanding of the change of
electronic structures above and below $T_{MI}$. Because Cu at A
site is in the Cu$^{1+}$ valence state (3d$^{10}$) \cite{Matsuno},
the Cu 3d band is fully filled. The band structure calculations
indicate that the Cu 3d band locates at about 3 eV below
E$_F$.\cite{Oda} Thus the MIT and the accompanied change is mainly
due to the variation of the electronic states of 5d transitional
metal Ir. Due to the crystal field and the hybridization between
Ir 5d and S 3p orbitals, the splitting of the $e_g$ and $t_{2g}$
bands of Ir 5d electrons is fairly large.\cite{Note2} As a result,
a low-spin state of Ir 5d electrons is favored. The Ir $e_g$ band
is empty and the states near Fermi level are mainly contributed by
the Ir $t_{2g}$ bands, but hybridized with S 3p orbitals.

Let us begin our discussion with the metallic phase in which \cis
has normal cubic spinel structure. There is only one equivalent
position for Ir in the structure with a valence state of
Ir$^{+3.5}$ and the Ir 5d band is partially filled. Band structure
calculation indicates that two bands arising from the
hybridization of Ir 5d$\epsilon$ (i.e. $t_{2g}$) and S 3p cross
the Fermi energy \cite{Oda}. These two bands lead to the Drude
responses in low frequencies. In this case, the electronic state
could be understood from the schematic picture of Fig. 3(a).

%
\begin{figure}[t]
\centerline{\includegraphics[width=2.9in]{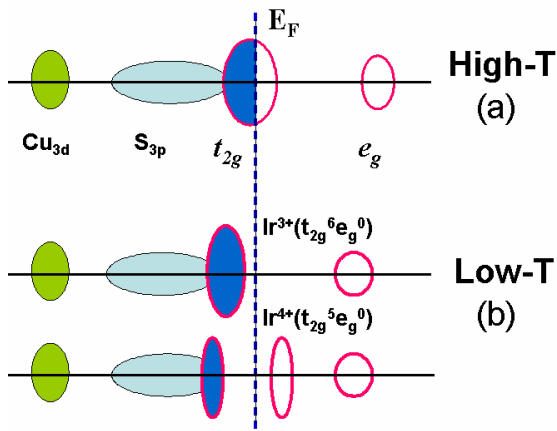}}%
\vspace*{-0.2cm}%
\caption{(Color online) Schematic diagrams of the electronic
states above and below the MIT temperature. The enclosed areas
above and below the horizontal lines represent the spin-up and
spin-down states. (a) Above $T_{MI}$, Ir$^{+3.5}$ has partially
filled $t_{2g}$ bands. (b) Below $T_{MI}$, the structure contains
only Ir$^{+3}$ and Ir$^{+4}$ octamers. Ir$^{3+}$ has fully filled
$t_{2g}$ band and completely empty $e_g$ band. The Ir$^{4+}$
should have a partially filled band in one of $t_{2g}$ orbitals.
However, the dimerization of Ir$^{4+}$ makes the band split into
two subbands.}%
\label{fig3}
\end{figure}

Upon entering the insulating state, a first-order structural phase
transition occurs and the lattice symmetry becomes triclinic
\cite{Radaelli}. Most remarkably, \cis undergoes a complex
charge-ordering transition. A group of 8 Ir$^{3+}$ or Ir$^{4+}$
octahedra forms a cluster called octamer, which can be viewed as
planar hexagonal rings with two additional octahedra attached to
the opposite sides. The Ir ions in the Ir$^{4+}$ octamers exhibits
drastic alternations of long and short Ir-Ir distances, whereas
the Ir-Ir distance in Ir$^{3+}$-octamers are uniform
\cite{Radaelli}. These structural features reveal important clues
for the understanding of the low-T optical spectra. Because of the
low-spin state of Ir ions, Ir$^{3+}$ has fully filled $t_{2g}$
bands and completely empty $e_g$ band. Therefore, the Ir$^{3+}$
(spin s=0) octamers are insulating. The Ir$^{4+}$ has
$t_{2g}^5e_g^0$ configuration (spin S=1/2) and one of the $t_{2g}$
orbitals is half filled. In principle, a material with a half
filled band should be metallic. However, the dimerization of
Ir$^{4+}$ ions splits this band into two subbands. The lower
subband is fully occupied while the upper subband is empty (Fig.
3b). Since the $\alpha$ peak appears only in the insulating phase,
it is reasonable to attribute the $\alpha$ peak to the transition
of electrons from the occupied Ir$^{3+}$ $t_{2g}$ or lower
Ir$^{4+}$ $t_{2g}$ to upper Ir$^{4+}$ $t_{2g}$ subband. Since the
transition of electrons from Ir$^{4+}$ site to Ir$^{4+}$ site
actually requires to overcome additional on-site Coulomb repulsion
energy, it is plausible that the lowest excitation is from
Ir$^{3+}$ $t_{2g}$ state to upper Ir$^{4+}$ $t_{2g}$ subband. This
is equivalent to say that the $\alpha$ peak is originated from the
inter-octamer hoppings. The $\beta$ peak comes from the transition
of electrons from the occupied Ir $t_{2g}$ to the empty Ir $e_g$
bands. Since the unoccupied Ir $e_g$ bands exist at temperature
higher than the MIT, the $\beta$ component is observable even in
the metallic phase. This is the reason why the
temperature-dependence of the $\beta$ peak is different from that
of the $\alpha$ one. The interband transition from Cu 3d to other
unoccupied state should appear at higher energies.

The optical data and the analysis provide a clear picture about
the change of electronic structures above and below the MIT
temperature. The high-T metallic state is due to the band
conduction of hybridized Ir $t_{2g}$ and S 3p electrons. In the
insulating state, the formation of the Ir$^{3+}$ and Ir$^{4+}$
octamers results in two different types of insulating clusters.
Ir$^{3+}$ octamers have fully occupied Ir $t_{2g}$ bands, whereas
Ir$^{4+}$ octamers produce two splitting subbands because of the
Ir$^{4+}$-Ir$^{4+}$ dimerization. Furthermore, the dimerized
Ir$^{4+}$ ions form a spin singlet. It suppresses the Pauli
paramagnetism of \cis and leads to the diamagnetic nature of the
insulating state \cite{Cao1}.

\begin{figure}[t]
\centerline{\includegraphics[width=2.9in]{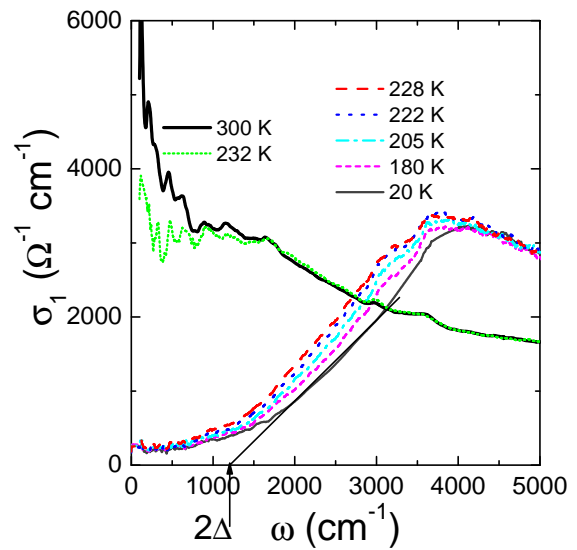}}%
\vspace*{-0.2cm}%
\caption{(Color online) The temperature dependence of the optical
conductivity in an expanded scale at low frequencies. A thin
straight line is extrapolation of the onset of the peak for the
estimate of the magnitude of the optical gap.}%
\label{fig4}
\end{figure}

The above discussion on the evolution of the electronic states is
consistent with a recent S \textit{K} and Ir \textit{$L_3$} x-ray
absorption study on \cis by Croft et al.\cite{Croft} where similar
redistribution of Ir 5d electronic states across the MIT has been
proposed. Apparently, the MIT of \cis is different to the Verway
transition in Fe$_3$O$_4$. Fe$_3$O$_4$ contains relatively narrow
3d band and the charge ordering is most likely caused by the
competition between the bandwidth and strong intersite Coulomb
repulsion \cite{note1,Cullen}. However, \cis is expected to have a
wide 5d band and weaker Coulomb repulsion. The MIT in \cis is due
to the reconstruction of Ir 5d bands associated with the
structural change.

Our result shows unambiguously that the MIT in \cis is directly
correlated with the structureal instability. It seems that this
structural instability is unique in the \cis family. Any
substitution to A-sites (e.g. Zn for Cu) \cite{Cao1,Cao2} or
B-sites (e.g. Rh for Ir) \cite{Matsumoto} or S-sites (e.g. Se for
S) \cite{Nagata2} will suppress the structural deformation and
drive the compound into metallic or superconducting in low
temperatures. The strong electron-phonon coupling is the most
probable mechanism responsible for the structural instability.
Further theoretical and experimental efforts to the understanding
of this mechanism are desired.

Fig. 4 shows the low-$\omega$ conductivity spectra in an expanded
scale. Below the MIT temperature, the optical conductivity
increases quickly above 1000 cm$^{-1}$. A rough estimation of the
optical gap could be obtained by extrapolating the linear
increasing part to the base line of $\sigma(\omega)$=0. This gives
the value of the optical gap 2$\Delta\sim$ 1200 \cm (~0.15 eV).
The gap magnitude ($\Delta$) is close to the activated gap values
estimated from several dc resistivity measurements
\cite{Nagata1,Furubayashi}. Matsuno et al. performed photoemission
measurements on \cis, but assigned a much smaller gap amplitude of
$\sim 20 meV$ in the insulating phase \cite{Matsuno}. However, by
looking at their spectral curves at 250 K and 30 K, we found that
the spectral edge actually shifts about 70 to 80 meV away from the
Fermi level, and seemed to be consistent with our experiment.
Additionally, we found that the energy gap changes very little as
the temperature increases from 10 K to 228 K. The sudden opening
of the energy gap below MIT is associated with the structureal
transition, and is a characteristic feature of the first-order
structural phase transition.

To conclude, optical conductivity spectra have been investigated
for single crystals of \cis. The metallic response at high
temperature is due to the band conduction of Ir $t_{2g}$
electrons, which are hybridized with S 3p electrons. The MIT in
\cis is caused by the reconstruction of Ir 5d bands associated
with the structural change. The formations of the Ir$^{3+}$ and
Ir$^{4+}$ octamers below T$_{MI}$ result in two different types of
insulating clusters. We attribute the $\alpha$ peak to the
transition of electrons from the occupied Ir$^{3+}$ $t_{2g}$ state
to upper Ir$^{4+}$ $t_{2g}$ subband created by the
spin-dimerization in the Ir$^{4+}$ octamers, and the $\beta$
component to the transition from the occupied Ir $t_{2g}$ to the
empty Ir $e_g$ bands.

This work is supported by National Science Foundation of China
(No. 10025418, 10104012, 10374109), the Knowledge Innovation
Project of Chinese Academy of Sciences.
%
%

\end{document}